# Magnetic phases of erbium orthochromite

Brajesh Tiwari#, M Krishna Surendra and M. S. Ramachandra Rao*

Nano Functional Materials Technology Centre, Materials Science Research Centre and Department of Physics, Indian Institute of Technology Madras, Chennai-600036, India

# Present address: Quantum Phenomenon and Application Division, National Physical Laboratory, New Delhi-110012, India.

* Corresponding authors: msrrao@iitm.ac.in

**Abstract:**

Erbium orthochromite, $ErCrO_3$, is a distorted-perovskite which has antiferromagnetic ground state below 10 K while $Cr^{3+}$ magnetic moments order at 133 K. The temperature dependence of magnetization is studied across different magnetic phases for $ErCrO_3$ and different magnetization isotherms are analyzed. In the presence of external magnetic field, polycrystalline $ErCrO_3$ develops weak ferromagnetism from antiferromagnetic ground state. These magnetic phase transitions are observed to be of first order which is justified by thermal hysteresis and Arrott-plots.

Keywords: Rare earth orthochromite, antiferromagnetism, first order magnetic phase transitions.

## 1. Introduction:

Rare earth orthochromites ($RCrO_3$) and orthoferrites ($RFeO_3$) which contain two magnetic ions ($3d^n$ and $4f^m$) have attracted research interest from several decades due to complex magnetic phases over temperature, pressure and magnetic field[1-4]. Recently these material systems received considerable attention in connection with magnetoelectric and multiferroic properties and their potential multifunctional applications[3,5,6,7]. The strong exchange interaction within the transition metal 3d, $Cr^{3+}(Fe^{3+})$- $Cr^{3+}(Fe^{3+})$ subsystems, is predominantly antiferromagnetic and usually orders at higher temperatures(several hundreds of Kelvin)than that of rare earth 4f,$R^{3+}$-$R^{3+}$subsystems,but are also less anisotropic compared to the rare earth ions. Hence rare earth magnetic ions can control the orientation of transition metal magnetic moments and give rise to complex magnetic structures. Such re-orientation transitions have profound effects on their magnetic, optical and elastic properties. It was shown by Hornreich *et al.* that the magnitude of rare earth-transition metal ion, $R^{3+}$- $Cr^{3+}$($Fe^{3+}$),coupling strength is large for orthochromites ($RCrO_3$) compared to orthoferrites ($RFeO_3$) because $Cr^{3+}$ does not have fourfold anisotropic terms [1]. This coupling plays a decisive role in determining the different magnetic phases in orthochromites due to spin reorientation transitions. These several important differences make orthochromites especially erbium orthochromite, $ErCrO_3$, more suitable for the study of magnetic phases depending on temperature, pressure and applied magnetic fields. $ErCrO_3$ belongs to the crystal space group $D_{2h}^{16}$ –Pbnm which contains four distorted perovskite units in the crystallographic cell[2,3]. Since $Er^{3+}$ ion has an electronic configuration of $4f^{11}$, the quantum number of total angular momentum J = 15/2 is a half integral number. Therefore, each multiplet can be split into J+1/2 Kramer's doublet in low symmetric crystals[6,8,9]. Due to interaction with $Cr^{3+}$ spins, this $Er^{3+}$ multiplet splitting reflects different effective fields in different magnetic phases which are given in table 1.

Table 1 Transformation properties of representation of space group Pbnm under generators $\underline{m}_x$, $\underline{m}_y$ and $m_z$ ($\underline{m}_x$, $m_y$ and $\underline{m}_z$. for Pnma)[2,3]. Considering chemical and magnetic unit cells, identical possible magnetic point groups and their compatibilities of Cr and Ln ions in $LnCrO_3$.

| Magnetic Symmetry Group | Compatible Spin Configurations (Magnetic Phases) | | | | |
|---|---|---|---|---|---|
| Point group: $d_{2h}$(mmm) | Pbnm setting | | | Pnma setting | |
| **Atom** | | Cr | Ln | Cr | Ln |
| mmm (mmm: $d_{2h}$) | $\Gamma_1$ | $A_xG_yC_z$ | $0'_x0'_yC'_z$ | $G_xC_yA_z$ | $0'_xC'_y0'_z$ |
| $\underline{mm}m$ (2/m:$c_{2h}$) | $\Gamma_2$ | $F_xC_yG_z$ | $F'_xC'_y0'_z$ | $C_xG_yF_z$ | $C'_x0'_yF'_z$ |
| | $\Gamma_3$ | $C_xF_yA_z$ | $C'_xF'_y0'_z$ | $F_xA_yC_z$ | $F'_x0'_yC'_z$ |
| | $\Gamma_4$ | $G_xA_yF_z$ | $0'_x0'_yF'_z$ | $A_xF_yG_z$ | $0'_xF'_y0'_z$ |
| $\underline{mmm}$ (222:$d_2$) | $\Gamma_5$ | | $G'_xA'_y0'_z$ | | $A'_x0'_yG'_z$ |
| $mm\underline{m}$ (mm2:$c_{2v}$) | $\Gamma_6$ | | $0'_x0'_yA'_z$ | | $A'_x0'_y0'_z$ |
| | $\Gamma_7$ | | $0'_x0'_yG'_z$ | | $G'_x0'_y0'_z$ |
| | $\Gamma_8$ | | $A'_xG'_y0'_z$ | | $G'_x0'_yA'_z$ |

Generators of the group Pbnm two glide planes; $m_x$: (x,y,z)→(1/2−x,1/2+y,z) and $m_y$:(x,y,z)→(1/2+x,1/2−y,1/2+z), and the mirror $m_z$: (x,y,z)→(x,y,1/2−z).

Different magnetic phases of $ErCrO_3$ have been studied using different experimental techniques such as neutron diffraction [3], magnetization measurements [4,5] and extensively by optical studies [6,8-11]. Below $T_N$ the spin structure of $Cr^{3+}$ ions in $ErCrO_3$ is $G_x$ and belongs to $\Gamma_4$ ($G_xA_yF_z$;$F'_z$) in Bertaut notations [2]. Spin reorientation transition $T_{SR}$ takes place abruptly at 10 K in absence of external magnetic field where $Cr^{3+}$ spins reorient in $G_y$ and belongs to $\Gamma_1$ ($A_xG_yC_z$;$C'_z$). This spin reorientation transition from $\Gamma_4$ ($G_xA_yF_z$;$F'_z$) to $\Gamma_1$ ($A_xG_yC_z$;$C'_z$), where the weak ferromagnetic moment disappear is a first-order phase transition based as symmetry arguments [12].It is therefore important to understand the magnetic response of $ErCrO_3$ in its different magnetic phases apart from phase transitions. The temperature dependence of magnetization was measured across different magnetic phases for $ErCrO_3$ and analyzed by different magnetization isotherms.

## 2. **Experimental details:**

The samples were prepared by conventional solid state reaction rout with nominal chemical composition of $ErCrO_3$ from starting materials of $Er_2O_3$ (99.9 % Alfa Aesar) and $Cr_2O_3$ (99.9 % Alfa Aesar). Before the final heat treatment at 1300 °C for 24 h stoichimetric powder of $Er_2O_3$ and $Cr_2O_3$ were mixed thoroughly in agate mortar and two intermediate calcinations were carried out at 600 °C and 900 °C for 12 h. The resulting dark green powder samples were used for structural and magnetic studies. The powder x-ray diffraction (XRD) data of the powder samples were collected using a PANalytical X'Pert Pro x-ray diffractometer with Cu Kα radiation under ambient conditions. Crystal structure refinements were carried out using General Structure Analysis System (GSAS) [10] for $D_{2h}^{16}$: Pnma (#62) and structural parameter were obtained. Magnetic measurements of $ErCrO_3$ were performed using vibrating sample magnetometer (VSM), an attachment in PPMS (Model 6000, Quantum Design, USA) in the temperature range of 5-300 K. Temperature dependent magnetization measurements were done as follows: zero-field cooled (ZFC) data collected while warming, field-cooled cooling (FCC) and field-cooled warming (FCW) procedures at an applied field. Magnetization (M) isotherms were recorded at different temperatures up to an applied magnetic field (H) of 7 kOe in vicinity of magnetic transitions.

## 3. Results and discussion:

### 3.1 Structural analysis:

The Rietveld refinement of $ErCrO_3$ X-ray powder diffraction data is shown in figure 1. The refinement was carried out using GSAS software for the orthorhombic crystal structure with space group *Pnma* (# 62)[13]. The difference-profile (Diff.) between the observed (Obs.) and calculated (Calc.) diffraction patterns is shown at the bottom of the plot. A good fit was obtained with *R* factors, w*R*p = 7.52 %, *Rp* = 5.23 %, and $\chi^2$ = 1.53. The lattice constants and volume of the unit cell are found to be $a$ = 5.512(1) Å, $b$ = 7.520(1) Å and $c$ = 5.226(1) Å and $V$ = 216.58(1) Å$^3$ respectively.

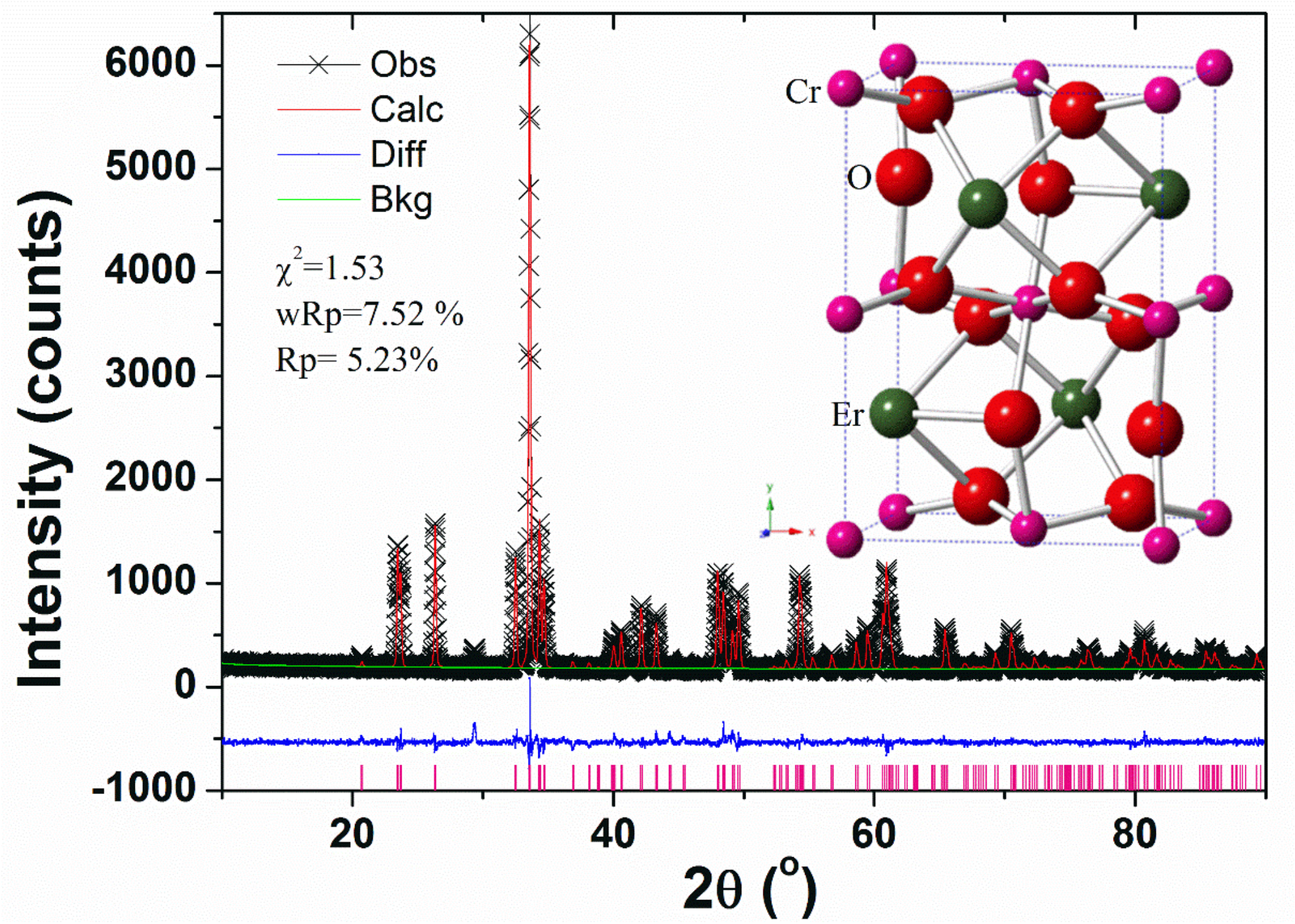

Figure. 1. X-ray diffraction pattern of $ErCrO_3$ which is Reitveld refined for model structure Pnma (Calc) with the observed pattern (Obs) to minimize the difference (Diff). The vertical bars represent the positions of Bragg reflections. Inset shows unit cell of $ErCrO_3$.

The inset in figure 1 shows the chemical unit cell of $ErCrO_3$ which has a total of 20 atoms (4 Er, 4 Cr and 12 O) per unit cell. Each chemical unit cell of $ErCrO_3$ has corner-linked octahedra $CrO_6$ with the centers occupied by centrosymmetric Cr ions (pink) with Wyckoff position 4b (0, 0, 1/2) while corner atoms of octahedra are oxygen ions (red) with two inequivalent positions, the apex oxygen (O1 ion) at 4c (-0.025, 0.250, 0.597) and planar oxygen (O2 ion) at 8d (0.714, -0.032, 0.307). Erbium ions (green) occupy the space among the octahedra at 4c (0.064, 0.25, 0.015). The distortion from ideal perovskite structure happens because of geometric tolerance factor of 0.903 as well as antiphase tilt of adjacent octahedra which in turn lead to Cr-O1-Cr bond angle ~ 144$^{o}$.

### 3.2 Magnetization study:

The temperature dependence of magnetization was measured across different magnetic phases for $ErCrO_3$ and analyzed by different magnetization isotherms. Figure 2 shows the magnetization curves as a function of temperature for $ErCrO_3$ recorded using an applied magnetic field of 100 Oe for different thermal cycles to understand the magnetic interactions; *first*, in zero-field sample is cooled down to 5 K and data recorded while warming (ZFC), *second*, data is recorded along with cooling in presence of external field (FCC) and *third*, field-cooled and data is recoded while warming (FCW). The temperature dependence of reciprocal of magnetic susceptibility $1/\chi$ is fitted to Curie-Weiss equation (red line), presented as right of y-axis. Inset of figure 2 shows magnified magnetization curves in the vicinity of antiferromagnetic ordering of $Cr^{3+}$ magnetic moments. Two distinct magnetic transitions can be observed; first at 133 K corresponds to $Cr^{3+}$ antiferromagnetic ordering and the second at 10 K related to the spin reorientation transition of $ErCrO_3$.

Above $T_N$ = 133 K in paramagnetic phase of $ErCrO_3$, $1/\chi$ vs. T curve follows the Curie-Weiss law and fitted as shown in figure 2. The estimated effective magnetic moment is $\mu_{eff} = 10.23\ \mu_B$ for $ErCrO_3$ which is close to the theoretical value $\mu_{th}(ErCrO_3) = 10.35\ \mu_B$ calculated from the free ion values 9.59 $\mu_B$ for $Er^{3+}$ and 3.87 $\mu_B$ for $Cr^{3+}$ (spin only values) moments added assuming their total randomness in paramagnetic phase i.e. $\mu_{th}(ErCrO_3) = \sqrt{\mu_{Er^{3+}}^2 + \mu_{Cr^{3+}}^2}\ \mu_B$. Below chromium ordering, it is also observed that magnetization gradually reaches a maximum at 19 K for $ErCrO_3$ showing paramagnetic 'Er' moments until ordering occurs at ~ 10 K. Asymptotically observed Weiss constant $\Theta$ = -29 K for $ErCrO_3$ is negative, indicating the predominance of antiferromagnetic interactions. The value of $|\Theta|/T_N$ for $ErCrO_3$ is 0.21 (<1) which differs from unity implying that the next-nearest neighbor couplings have considerable strength for the determination of ground state magnetic structure of $ErCrO_3$which is effectively antiferromagnetic.

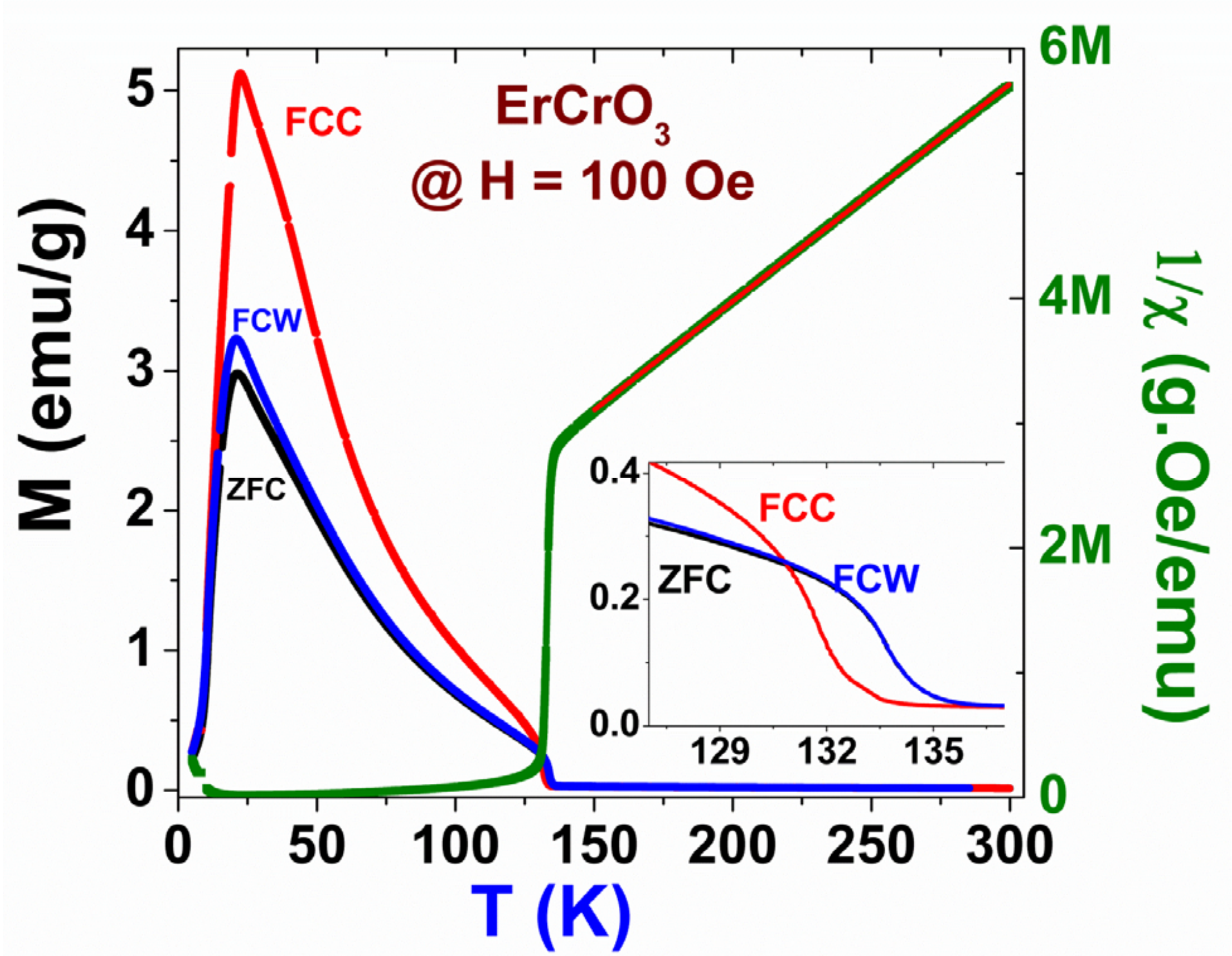

Figure 2 Magnetization of $ErCrO_3$ recorded under 100 Oe applied magnetic field for different thermal cycle; first, zero-field cooled down to 5 K and data recorded while warming (ZFC), second, field-cooled (100 Oe) and data is recorded along with cooling (FCC) and third, field-cooled (100 Oe) and data is recoded while warming (FCW). The temperature dependence of reciprocal of magnetic susceptibility $1/\chi$ is fitted to Curie-Weiss equation (red line), presented on the right hand side of y-axis. Inset: Magnetization curves are magnified in the vicinity of antiferromagnetic ordering of $Cr^{3+}$ magnetic moments.

Magnetization curves show a thermal hysteresis in the vicinity of antiferromagnetic ordering of $Cr^{3+}$ magnetic moments as shown as inset of figure 2 (a magnified view of magnetization curves). The magnetization onset for ZFC and FCW curves is 133.7 K while FCC shows at 131.6 K. A difference of 4.6 K is observed when magnetization is recorded during cooling of $ErCrO_3$ in presence of magnetic field (FCC) compared to data recorded during warming (FCW and ZFC) which will be explained latter which is significant compared to $HoCrO_3$ and $YCrO_3$[14].

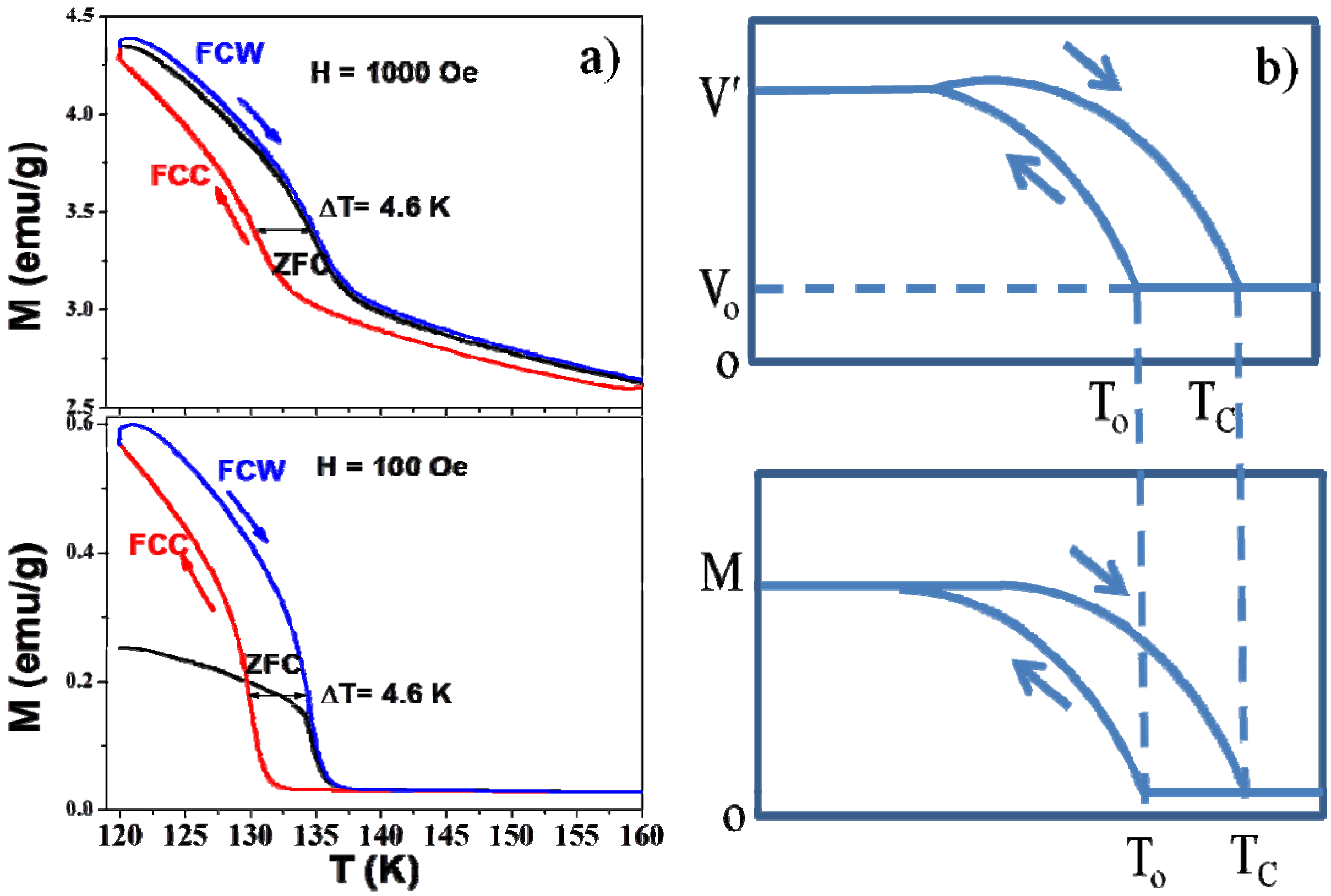

Figure 3 a) ZFC (black) FCC (red) and FCW (blue) magnetization as a function of temperature with two different applied magnetic fields 100 Oe (lower panel) and 1000 Oe (upper panel). There is a thermal hysteresis between FCC and FCW with a clear temperature difference of 4.6 K. b) Schematic of thermal hysteresis in magnetization due to change in unit cell volume from free (V') to the clamped ($V_0$) in presence of magnetic field in its magnetic state (see the text).

Figure 3 (a) shows a clear thermal hysteresis in the magnetization curves for applied magnetic fields 100 Oe (lower panel) and 1000 Oe (upper panel) with a temperature difference of 4.6 K. In La doped $GdCrO_3$ a similar thermal hysteresis was observed by Sharma et.al[15] concluding it to be first order phase transition between different magnetic phases. This type of thermal hysteresis in magnetization measurements can be explained by considering the possibility that the lattice is deformable and a spontaneous lattice deformation of lattice occurs in the magnetically ordered state[16]. Though there are several other origins of this type of thermal hysteresis. For example in doped Lanthanum manganites systems which show thermal hysteresis due to first order ferromagnetic to antiferromagnetic in conjugation to metal to insulator transition in orbital and/or charge order systems.[17] [18] [19] The exchange interaction that gives rise to magnetically ordered state and also determine the transition temperature is a strong function of interatomic spacing. Figure 3 (b) shows the schematic to explain the phenomenon that leads to thermal hysteresis in magnetization. In the high temperature state and above Curie temperature the lattice volume is $V_o$ and the transition temperature is $T_C$, if the sample is cooled across the transition temperature then the volume due to lattice distortion is V' and corresponding transition temperature is $T_o$. In the cooled state if the sample is warmed then the volume is V' and the corresponding transition temperature $T_o$ is seen. This can be understood based on the fact that the if the sample is cooled to the lowest temperature in the presence (or absence) of field which can distort the lattice (or *free system*) at the transition temperature ($T_C$ or $T_N$) and if the sample is warmed from its lowest temperature (in figure 2 down to 5 K and figure 3 (a) down to 120 K) then the free energy may be lowered in the direction of increasing transition temperature ($T_C$ or $T_N$) as in the case when magnetization is recorded while warming, i.e. FCW (or ZFC). Now if magnetization data while warming (FCW or ZFC) is compared with the magnetization curve recorded while cooling (FCC) the sample shows its intrinsic (or *clamped system*) lattice volume ($V_0$) and a lower magnetic ordering temperature ($T_0$).

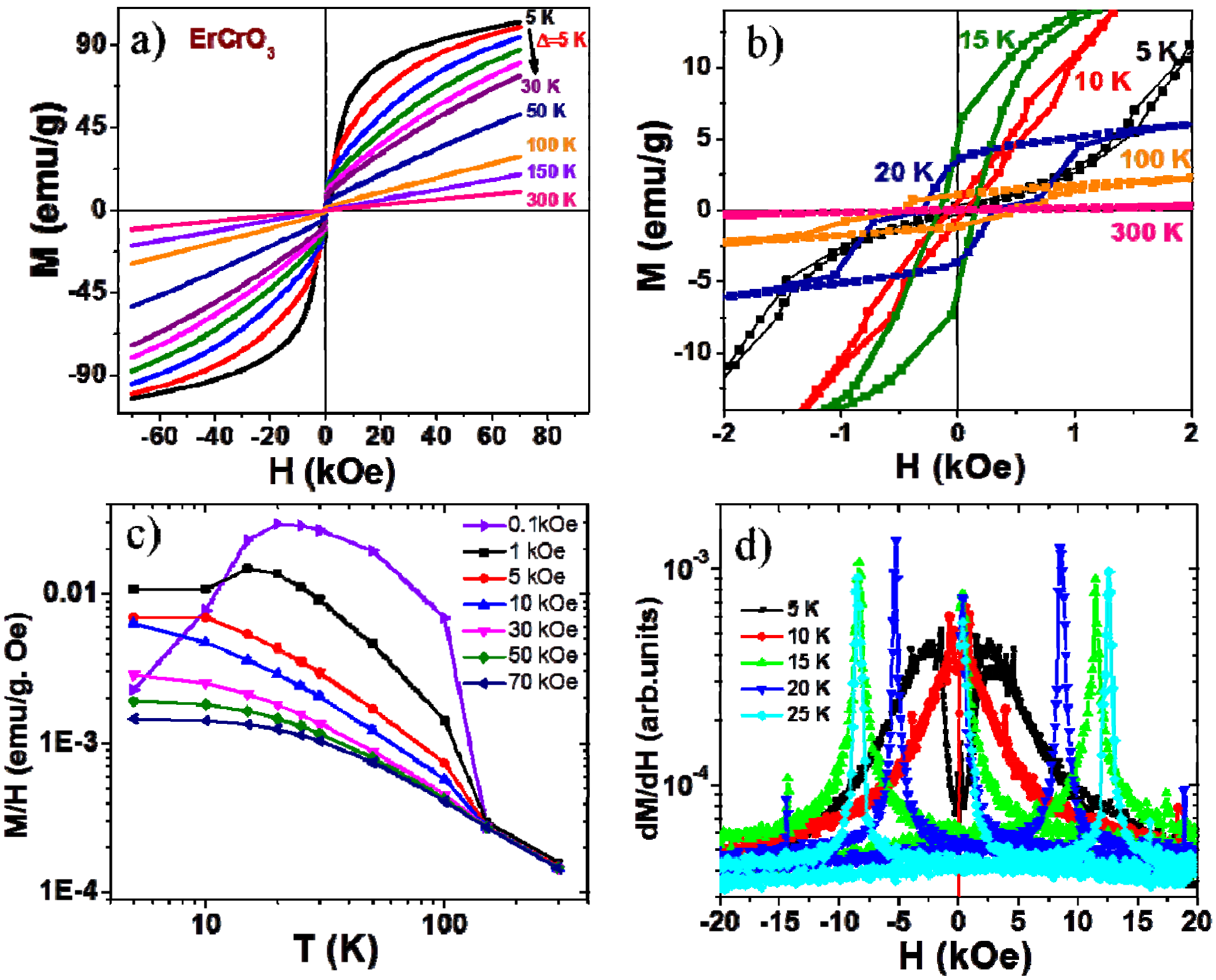

Figure 4 Magnetization measurements of $ErCrO_3$ at different temperatures. a) Magnetization curves at different temperatures in the range 5 K to 300 K with an applied magnetic field up to 70 kOe and b) magnified view of magnetization curves at low fields up to 2 kOe (all curves are not shown for clarity). c) Magnetic susceptibility (M/H) as a function of temperature at different applied magnetic fields as derived from magnetization curves. d) First derivative of dM/dH magnetization isotherms at 5, 10, 15, 20 and 25 K (for clarity other isotherms are not shown) show sharp maxima at different fields indicating a clear change in magnetization behavior.

Further to understand the different magnetic phases and nature of magnetic transitions of $ErCrO_3$, magnetization isotherms over the temperature range 5 K to 300 K were recorded and shown in figure 4. Magnetization curves up to applied magnetic fields of 70 kOe as shown in figure 4 (a) indicate a sign of magnetization saturation only at 5 K while others isotherms above the spin reorientation transition of 10 K do not saturate. Magnetization curves for low applied magnetic fields (-2 kOe to 2 kOe) are shown in figure 4 (b). Magnetic susceptibilities (M/H) over the temperature range 5 - 300 K are shown in log-log scale in figure 4 (c) which was obtained from magnetization isotherms from figure 4 (a). At low magnetic fields (0.1 kOe and 1.0 kOe) spin reorientation transition ($T_{SR}$ = 10 K) can be observed which is similar to magnetization as a function of temperature (figure 2), but for high applied magnetic fields, this spin reorientation transition gets suppressed as can be seen from figure 4 (c) retaining weak ferromagnetic nature of $ErCrO_3$. The applied magnetic field which suppresses the spin reorientation transition can be determine by the first derivative (dM/dH) of magnetization isotherms, as shown in figure 4 (d) for temperatures close to $T_{SR}$. For temperatures above spin reorientation ($T > T_{SR}$), dM/dH shows two sharp maxima one in the magnetic fields 5-12 kOe while second around 16 kOe, indicating two magnetic behavior/phases change which are weak ferromagnetic phases of $ErCrO_3$, $\Gamma_2(C_xG_yF_z; C_x^{'}F_z^{'})$ and $\Gamma_4(A_xF_yG_z; F_y^{'})$. Also it is important to note that the magnetic susceptibility dM/dH, at 5 K, shows minimum value in absence of external field in contrast to other higher temperatures which justifies the fact that ground state of $ErCrO_3$ is antiferromagnetic $\Gamma_1(G_xC_yA_z; C_y^{'})$.

Magnetization curve at 5 K does not show any loop for low field opening indicating perfect antiferromagnetic phase $\Gamma_1(G_xC_yA_z; C_y^{'})$ in the absence of strong magnetic fields for $ErCrO_3$ in Bertaut notation which are given in table 1. As the applied magnetic fields increase, the magnetization starts increasing and tending towards saturation which corresponds to weak ferromagnetic phases $\Gamma_2(C_xG_yF_z; C_x^{'}F_z^{'})$ when applied magnetic field is above 1 kOe along crystallographic z-axis and $\Gamma_4(A_xF_yG_z; F_y^{'})$ and when the applied magnetic field is above 10 kOe along crystallographic y-axis. These magnetic phases are well studied using optical absorption by Hasson *et.al.* and Toyokawa *et. al*[8,10]. These field induced transitions are first order phase transition. Even in the absence of magnetic field along y-axis $Cr^{3+}$ magnetic moments undergo a spin reorientation-type transition around 10 K from weak ferromagnetic $\Gamma_4(A_xF_yG_z; F_y^{'})$ phase to $\Gamma_1(G_xC_yA_z; C_y^{'})$ in which the ferromagnetic moments vanish.

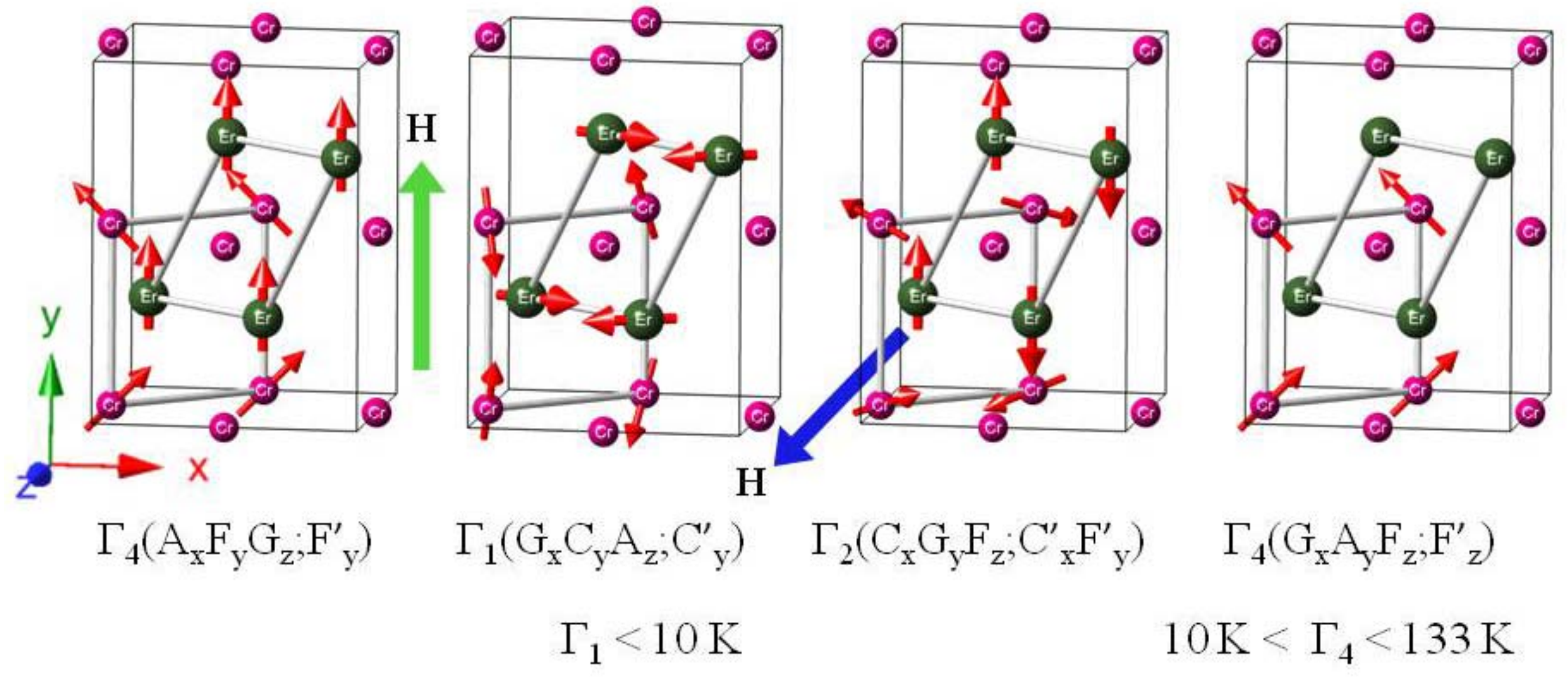

Figure 5 Different magnetic phases of $ErCrO_3$ below $Cr^{3+}$ magnetic ordering (133 K) and other magnetic phases depending on temperature and direction of external magnetic field **H** (O atoms are not shown for clarity).

Since in the present study, polycrystalline $ErCrO_3$ is used, apart from 5 K magnetization isotherm at low magnetic field which is perfect antiferromagnetic with $\Gamma_1(G_x C_y A_z; C_y^{'})$ as shown in figure 5, however magnetization isotherms between $T_{SR}$ and $T_N$ show the magnetic phase of$\Gamma_4(A_x F_y G_z; F_y^{'})$ even in absence of magnetic field. In $\Gamma_1(G_x C_y A_z; C_y^{'})$ phase below $T_{SR}$ ~ 10 K, if external magnetic field is applied parallel crystallographic z-axis above some critical value $H_{c//z}$ the spin of $Cr^{3+}$ reorientation themselves from $G_x$ to $G_y$ and induces weak ferromagnetism $F_z$ in z-direction which is represented as $\Gamma_2(C_x G_y F_z; C_x^{'} F_z^{'})$. Similarly either by increasing temperature above $T_{SR}$ or by applying external magnetic field parallel to y-axis above a critical value $H_{c//y}$ results in $\Gamma_4(A_x F_y G_z; F_y^{'})$ which have ferromagnetic component $F_y$ in y-direction, see figure 5. These weak ferromagnetic phases of $ErCrO_3$up on application of external fields $\Gamma_2(C_x G_y F_z; C_x^{'} F_z^{'})$and$\Gamma_4(A_x F_y G_z; F_y^{'})$ along two different directions as observed as two magnetization behavior in figure 5. Transitions between two magnetic phases of $ErCrO_3$ either by magnetic field or upon temperature are first order which can be justified by Arrot plots which is based on phenomenological Ginzberg-Landau theory in vicinity to these magnetic transitions [12,20,21].

Ginzburg-Landau formulation which includes the magnetostatic field energy (*MH),* where $M$ is experimentally observed specific magnetization as an order parameter and H is the applied magnetic field, the thermodynamic potential is given by $\Phi = \Phi_0 + \frac{1}{2}\alpha M^2 + \frac{1}{4}\beta M^4 - MH$,where α, β are temperature dependent constants. In equilibrium $\partial\Phi/_{\partial M} = 0$ the expression reduces to $H/_M = \beta M^2 + \alpha$. Arrott plot, isotherms of $M^2$against $H/_M$should be straight line in high applied magnetic field [21]. For second order phase transitions, β should positive while negative for first order magnetic phase transition according to the criteria proposed by Banerjee *et al.*[20]. In order to confirm the two magnetic transitions in $ErCrO_3$, antiferromagnetic Cr with Neel temperature $T_N$ = 133 K and spin reorientation transition at $T_{SR}$ = 10 K which show thermal hysteresis (figure 2) are first order magnetic transitions or not, first quadrant magnetization isotherms in vicinity of these transitions are recorded.

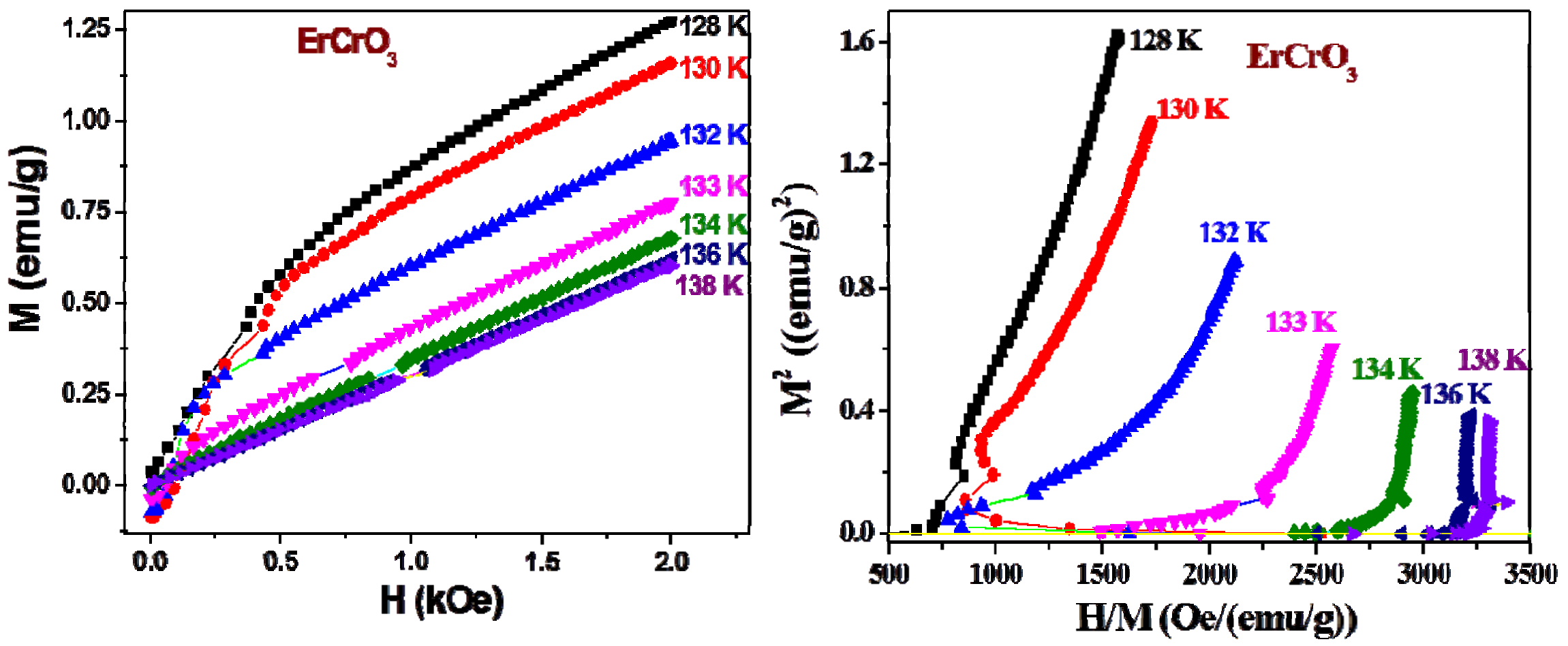

Figure 6 a) Magnetization isotherms close to Cr ordering temperature and b) corresponding Arrot's plots $M^2$~H/M at different temperatures indicates first order transition as slope just below transition is negative.

Magnetization isotherms at different temperatures between 128 K to 138 K are shown in figure 6 for $ErCrO_3$. A linear increase in the magnetization above 133 K (134, 136 and 138 K) over the magnetic range (0-2 kOe) is clear signature paramagnetic phase of material as shown in figure 6 (a). Below $T_N$=133 K a nonlinear behavior indicates the antiferromagnetic phase of $ErCrO_3$. Arrot plots as presented in figure 6 (b) show the slopes of $M^2$ vs. $H/_M$ curves negative below the $T_N$ (133 K) means negative β which supports the observed thermal hysteresis (about 4.6 K) in FCC and FCW magnetization measurements as shown in figure 2 and figure 3.

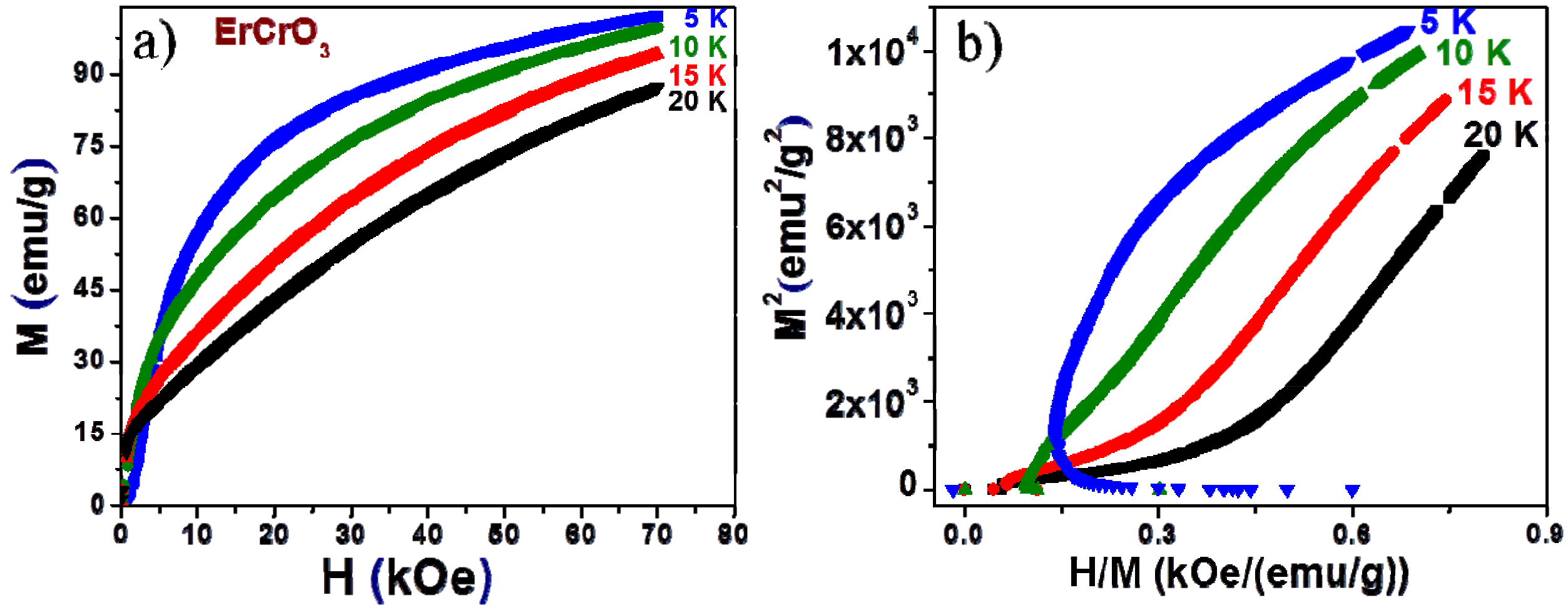

Figure 7 a) First coordinate magnetization isotherms in vicinity of spin reorientation transition temperature and b) corresponding Arrot's plots which show a clear negative slope at 5 K for $ErCrO_3$.

As it is observed from magnetization measurements there is a spin reorientation transition ($T_{SR}$= 10 K) from high temperature $\Gamma_4(A_xF_yG_z; F_y^{'})$ to low temperature $\Gamma_1(G_xC_yA_z; C_y^{'})$ in the absence of external magnetic fields. Transition between these two magnetic phases should genuinely show first order transition though not necessary. To confirm, first quadrant magnetization isotherms were recorded as shown in figure 7 (a) and the corresponding Arrot plots are shown in figure 7 (b). At spin reorientation transition i.e. $T_{SR}$ =10 K $M^2$ vs. H/M curve is linear indicating the transition whereas at 5 K (< $T_{SR}$) shows a clear negative slope for low magnetic fields indicating this transition to be first order.

For $ErCrO_3$ two first order magnetic transitions can be observed. A canted antiferromagnetic or weak ferromagnetic order of phase $\Gamma_4(A_xF_yG_z; F_y^{'})$of $Cr^{3+}$ ions occurs at $T_N$= 133 K while presence of second magnetic ion $Er^{3+}$ which has aspherical charge distribution with non-zero net magnetic moment, reorient the $Cr^{3+}$ such that a new magnetic phase $\Gamma_1(G_xC_yA_z; C_y^{'})$ appears at $T_{SR}$ = 10 which makes $ErCrO_3$ perfect antiferromagnetic (ferrimagnetic). These transitions proved to be of first order whose origin lies in the dependence of exchange energy as a strong function of inter-atomic spacing and bond strength. At absolute zero the distortion of lattice favors increasing transition temperature as we observed for $ErCrO_3$in which the onset of magnetic ordering has been found to increase by ~4.6 K in FCW compared to FCC measurement cycle.

**Conclusion:**

Erbium orthochromite in its distorted-perovskite structure is synthesized and magnetic properties are studied. We observed a thermal hysteresis of 4.6 K in magnetization and is independent of applied magnetic field strength (upto 1000 Oe) which indicates transition type is first order. Below 10 K, the complex magnetic phases indicate the significance of Er-Cr coupling in $ErCrO_3$. At 5 K, with external magnetic field the antiferromagnetic ground state $\Gamma_1(G_x C_y A_z; C_y^{'})$ changes to weak ferromagnetic $\Gamma_4(A_x F_y G_z; F_y^{'})$ and $\Gamma_2(C_x G_y F_z; C_x^{'} F_z^{'})$ phases.

**Acknowledgments:**

The authors would acknowledge the Department of Science and Technology (DST) of India for the financial support (grant No. SR/NM/NAT-02/2005).